\documentclass{cicp}
\usepackage{graphicx}

\usepackage{amsmath}
\usepackage{amssymb}
\usepackage{amsbsy}
\usepackage{textcomp}
\usepackage{MnSymbol}

\usepackage{color}
\definecolor{orange}{rgb}{1,0.5,0}

\def\<{\langle} \def\>{\rangle}

\begin{document}

\title{Efficient Computation of Instantons for Multi-Dimensional
  Turbulent Flows with Large Scale Forcing}

\author[T. Grafke et.~al.]{
   Tobias Grafke\affil{1}\comma\corrauth,
   Rainer Grauer\affil{2}, and
   Stephan Schindel\affil{2}}
\address{\affilnum{1}\ 
  Department of Physics of Complex Systems, 
  Weizmann Institute of Science, 
  Rehovot 76100, Israel. \\
  \affilnum{2}\ 
  Theoretische Physik I, 
  Ruhr-Universit\"at Bochum,
  Universit\"atsstr. 150, 
  D44780 Bochum (Germany). \\
}

\email{{\tt tobias.grafke@weizmann.ac.il} (T.~Grafke)}

\date{\today}

\begin{abstract}
  Extreme events play a crucial role in fluid turbulence. Inspired by
  methods from field theory, these extreme events, their evolution and
  probability can be computed with help of the instanton formalism as
  minimizers of a suitable action functional. Due to the high number
  of degrees of freedom in multi-dimensional fluid flows, traditional
  global minimization techniques quickly become prohibitive in their
  memory requirements. We outline a novel method for finding the
  minimizing trajectory in a wide class of problems that typically
  occurs in turbulence setups, where the underlying dynamical system
  is a non-gradient, non-linear partial differential equation, and the
  forcing is restricted to a limited length scale. We demonstrate the
  efficiency of the algorithm in terms of performance and memory by
  computing high resolution instanton field configurations
  corresponding to viscous shocks for 1D and 2D compressible flows.
\end{abstract}

\pac{47.11.-j, 47.27.ef, 05.10.-a, 47.40.-x}

\maketitle

\section{Introduction}

Systems in nature are almost always subject to noise. Even though
these random perturbations often are small in amplitude, they
nevertheless have drastic consequences on the behavior of the system
as a whole by facilitating rare but extreme excursions of the
dynamics. Many processes in biology, chemistry, physics and economics,
including phase transitions \cite{elgart-kamenev:2006}, ocean dynamics
\cite{schmeits-djikstra:2001}, rates of chemical reactions
\cite{kramers:1940}, genetic switches \cite{assaf-roberts-etal:2011}
and option pricing in finance \cite{barro:2006,backus-chernov-martin:2011}
are caused by rare extreme events.

In the small noise limit, these extreme events are in fact very
predictable. The arising most probable transition trajectory of the
rare event in turn allows for predictions regarding both the evolution
of the rare event and its probability. In the context of the
Martin-Siggia-Rose/Jansen/de Dominicis formalism
\cite{martin-siggia-rose:1973, janssen:1976, dedominicis:1976} these
field configurations are termed \emph{instantons}. They correspond to
the minimizers of the rate function in the Freidlin-Wentzell theory of
large deviations \cite{freidlin-wentzell:2012, dembo-zeitouni:2010}.

A number of numerical algorithms have been devised to compute these
field configurations. Some of them, like the nudged elastic band
method \cite{henkelman-jonsson:2000} or the string method
\cite{e-ren-vandeneijnden:2002}, are only applicable to the important
sub-class of gradient systems, while others, most notably the minimum
action method \cite{e-ren-vandeneijnden:2004} and variants thereof
\cite{heymann-vandeneijnden:2008, zhou-ren-e:2008}, are able to find
the instanton configuration for more general cases. All have in common
that they solve the problem \emph{globally}, by discretizing the
trajectory along the physical time and applying global operations on
its entirety. For PDE systems with an infinite number of degrees of
freedom, in particular in higher dimensions, such as arising in
turbulent fluids, the memory requirements of these algorithms quickly
become prohibitive.

Many questions in fluid dynamics, such as shock formation in
compressible flows \cite{grafke-grauer-schaefer:2013} or the
generation of rogue waves \cite{onorato-residor-etal:2013}, allow an
alternative to the global formulation due to the nature of their mixed
initial/final boundary conditions. Here, it is feasible to iteratively
solve the equations of motion (the \emph{instanton equations}) of the
underlying Hamiltonian system instead
\cite{chernykh-stepanov:2001}. The boundary conditions are propagated
throughout the domain, which opens up possibilities to avoid saving
the field configuration at every instance in time. In particular, a
drastic reduction in memory is possible by combining a number of
techniques: (a) considering the ``geometric'' reparametrization of the
trajectory instead of parametrization by physical time, (b) recursive
storage of transition states, inspired by multigrid techniques, (c)
exploiting the compactness of the support of the force correlation in
turbulence setups, and (d) further memory reductions through wavelet
compression. The detailed presentation of the resulting algorithm
constitutes the core of this work. We illustrate its effectiveness by
applying it to 1D and 2D compressible turbulence. In fact we
demonstrate that the combined optimizations reduce the memory
footprint enough to fit optimization problems with $N=10^{10}$ degrees
of freedom on a single graphics card. It therefore becomes feasible to
solve the numerical problem on graphics processing units (GPUs)
instead of their host machines at a considerable gain in runtime
performance. In consequence, all algorithms presented in this paper
are implemented on GPUs using the CUDA framework
\cite{cuda:2014}. Furthermore, due to its memory efficiency, the
scheme in principle allows attacking the important problem of the
computation of instanton configurations for the 3D incompressible
Navier-Stokes equation, which would yield scaling predictions for
turbulent statistics.

This paper is organized as follows: We first establish the instanton
formalism and the associated minimization problem in
section~\ref{sec:2}. We introduce the Martin-Siggia-Rose/Janssen/de
Dominicis formalism and present a beneficial modification of the
action functional in terms of a geometric reparametrization. In
section~\ref{sec:3}, we outline the composition of our proposed
algorithm. This includes a discussion of the recursive integration of
the mixed initial/final value problem posed by the instanton equations
and its combination with the projection of the auxiliary field
variable. Section~\ref{sec:4} presents applications of the proposed
algorithm to the problem of prototypical shock configurations of one-
and two-dimensional compressible fluids. We demonstrate the efficiency
in memory of the algorithm by computing the minimizer in a high
degrees of freedom setup and discuss the computational overhead of the
optimizations in terms of performance. We conclude in
section~\ref{sec:5} with a discussion of the results, present possible
future applications as well as comment on their potential impact on
the understanding of turbulent fluids.

\section{Instantons formalism for stochastic partial differential equations}
\label{sec:2}

Consider a system of $n$ partial differential equations driven by a
Gaussian noise,
\begin{equation}
  \label{eq:spde}
  \dot u = b[u] + \eta(x,t)
\end{equation}
in $d$-dimensional domain, i.e. the field $u$ is a function
$u(x,t):\mathbb{R}^d\times[-T,0] \rightarrow \mathbb{R}^n$. The drift
$b[u]$ is some possibly nonlinear operator and the Gaussian noise
$\eta$ has a finite correlation length $L$ in space and is
white-in-time,
\begin{equation}
 \label{eq:correlation}
 \< \eta(x,t) \eta(x+r,t+s) \> = \chi(r)\delta(s).
\end{equation}
We restrict ourselves to the case of a correlation function $\chi(r)$
with compact support in Fourier space,
\begin{equation}
 \label{eq:space_corr}
 \hat\chi(k) = f(k) \theta(k_c-|k|),
\end{equation}
for an arbitrary shape $f(k)$, ${\theta}$ denoting the Heaviside step
function. Thus, the driving occurs only at scales larger than a
cut-off scale $L \sim 1/k_c$. This setup appears naturally in
turbulent fluid systems with a direct cascade, such as 3D
incompressible Navier-Stokes turbulence, MHD turbulence or Burgers
turbulence: Energy is inserted at large scales and transported to
small scales by the non-linear term, where it is dissipated by
viscosity. Note that the opposite choice of small scale forcing, which
is necessary for configurations with an inverse cascade, is equally
viable despite being not of the form~\eqref{eq:space_corr}, as the
only necessary condition is a compact support of the forcing
correlation.

The Martin-Siggia-Rose/Janssen/de Dominicis path integral formalism
allows us to formally write down the expectation of any observable of
the field variables, $\mathcal{O}[u]$, under the noise $\eta$ as
\begin{equation}
  \label{eq:pathintegral}
  \langle \mathcal{O}[u] \rangle_\eta \propto \int Du\, \int D(ip)
  \mathcal{O}[u] \exp (-I_T[u,p])
\end{equation}
with the \emph{action functional} or \emph{response functional}
\begin{equation}\label{eq:action}
  I_T[u,p] = \int_{-T}^0 \left(\langle p,\dot u - b[u] \rangle - \tfrac12 \langle p, \chi p \rangle \right) dt\,,
\end{equation}
introducing the \emph{auxiliary field} $p$. Here, $\langle \cdot,
\cdot \rangle$ is the $L_2(\mathbb{R}^n)$-scalar product. Depending on
the exact form of the observable at hand, it is then possible to
estimate the expectation by approximating the path integral
\eqref{eq:pathintegral}. For the case of transition possibilities
between two known states $u_-$ and $u_+$, this is usually done by a
saddle point approximation, which amounts to finding the minimizers of
the action functional \eqref{eq:action}, which are also termed
\emph{instantons} and correspond to the classical trajectories under
the action functional $I_T[u,p]$. In fact, in the limit of extremely
rare events, the saddle point approximation becomes exact, as
rigorously derived in the Freidlin-Wentzell theory of large deviations
\cite{freidlin-wentzell:2012}. To numerically find the minimizer of
\eqref{eq:action}, one discretizes the action and subsequently employs
global minimization techniques to solve the variational problem
\begin{equation}
  \label{eq:minimization}
  \inf_{T>0} \, \inf_{u} \, I_T[u,p]
\end{equation}
with appropriate boundary conditions $u(x,t=-T)=u_-(x)$,
$u(x,t=0)=u_+(x)$ (the auxiliary field is usually computed explicitly
as $p=\chi^{-1}(\dot u-b[u])$ in this setup). This approach is
commonly taken in practice, most notably by the string method
\cite{e-ren-vandeneijnden:2002} for gradient systems, and by the
minimum action method \cite{e-ren-vandeneijnden:2004} and variants
thereof \cite{zhou-ren-e:2008, vandeneijnden-heymann:2008} in more
general setups.

For a wide class of questions the form of the observable
$\mathcal{O}[u]$ allows for an efficient alternative approach: Suppose
we only want to measure the observable at the final time $t=0$ (or any
other distinct point in time), and furthermore only measure a single
degree of freedom in the final field configuration. This could, for
example, be the ocean surface elevation at a single point in a rogue
wave setup, or an extreme velocity gradient for Burgers shock
formation (see below). Note that in contrast to the computation of
transition probabilities, the exact form of the final condition of the
field $u$ is not prescribed, but instead is recovered as part of the
solution. This allows to recover the exact \emph{form} in space and
time of the extreme ocean surface wave form or extreme shock structure,
respectively, from the formalism.
In the outlined setup, the observable takes the form 
\begin{equation*}
  \mathcal{O}[u] = \delta(F(u(x,t=0))-a)
\end{equation*}
for a scalar functional $F(u)$. For example, we choose $F(u)=u_x
\delta(x)$ to measure the probability of a gradient $a$ in the origin,
or $F(u)=u \delta(x)$ to measure a surface elevation of $a$ in the
origin. In general, we can rewrite this observable as a Fourier
integral,
\begin{equation*}
  \label{eq:observable}
  \mathcal{O}[u] = \delta(F(u(x,t=0))-a) = \int d(i\lambda) \exp\big(\int_{-T}^0 \lambda (F[u]-a) \delta(t) \big)
\end{equation*}
and insert it into the path integral formulation \eqref{eq:pathintegral},
\begin{equation}
  \langle \mathcal{O}[u] \rangle_\eta \propto \int Du\, \int D(ip)
  \int d(i\lambda) \,\exp \big[-\underbrace{\big(I_T[u,p] - \lambda (F[u]-a)\big)}_{S_T[u,p]} \big]\,,
\end{equation}
where we label the modified action functional in the exponent of the
path integral as $S_T[u,p]$. Similar to the case of transition
probabilities, we want to estimate the expectation by employing a
saddle point approximation and computing the instanton field
configuration. Instead of globally minimizing the action functional we
resort to solving the equations of motion or \emph{instanton
  equations} for the associated Hamiltonian system instead
\cite{chernykh-stepanov:2001, grafke-grauer-schaefer:2013}. These
follow from the realization that at the saddle point the variation of
the functional $S_T[u,p]$ with respect to the fields $u,p$ vanishes,
leading to
\begin{subequations}
  \label{eq:instanton_t}
  \begin{align}
    \dot u &= b[u] + \chi p \label{eq:instanton_t1}\\ 
    \dot p &= -(\nabla b[u])^T p + \lambda\nabla F[u]
              \delta(t)\,, \label{eq:instanton_t2}
  \end{align}
\end{subequations}
where $\nabla$ is meant in the functional sense with respect to the
field variable. Now, the second term of the auxiliary equation,
$\lambda\nabla F[u] \delta(t)$, which incorporates the observable,
acts only on the final time $t=0$. It can therefore be understood as a
\emph{final condition} for the auxiliary field
\cite{gurarie-migdal:1996, balkovsky-falkovich-etal:1997}. On the
other hand, the final configuration of the field $u$ is not known. For
this reason, global minimization techniques are harder to implement in
this case. Though in principle a modification of the variational
formulation~\eqref{eq:minimization} is possible in terms of
e.g. penalty methods which enforce the constraint on the final field
configuration, integrating the equations of motions is a much more
natural approach: The boundary constraints are automatically fulfilled
when propagating the initial configuration of $u$ forward in time via
equation~\eqref{eq:instanton_t1}, while propagating the final
conditions of $p$ backwards in time via
equation~\eqref{eq:instanton_t2}. The final field configuration
$u(x,t=0)$ adheres the constraints of the observable and at the same
time yields the most probable composition of the unconstrained degrees
of freedom.
Note also, that equation~\eqref{eq:instanton_t1} resembles the
original stochastic system \eqref{eq:spde}, with the random noise
$\eta$ being replaced by the convolution $\chi p$. A solution of the
instanton equations~\eqref{eq:instanton_t} therefore not only allows
for a prediction of the evolution of the field leading up to the
extreme event at $t=0$, but additionally yields the corresponding
force.

\subsection{Geometric action functional and arc-length reparametrization}

We want to consider the case $T \rightarrow \infty$, i.e. the infinite
time minimizer. This case arises in general when asking for the
emergence of an extreme event out of a stationary state, which in
terms of large deviations corresponds to a diffusive exit from a
stable fixed point of the unperturbed dynamics. From a numerical point
of view, the limit $T \rightarrow \infty$ poses the obvious problem of
how to discretize an infinite time interval appropriately. 

For minimization problems of the form~\eqref{eq:minimization}, the
form of the instanton configuration is independent of the
parametrization of the trajectory. It is therefore possible to
reformulate the variational problem into a minimization on the space
of arc-length parametrized curves instead. The modified action
functional,
\begin{equation}
  \label{eq:gmam_action}
  I[u,\dot u] = \int_0^1 \left( \|\dot u\|_{\chi} \|b[u]\|_{\chi} -
  \llangle \dot u,b[u]\rrangle_{\chi} \right) ds,
\end{equation}
is shown to be independent of the parametrization
\cite{heymann-vandeneijnden:2008}, such that the search space is
restricted to trajectories of unit length. The metric is induced by
the correlation matrix $\chi$ of the problem via
\begin{equation}
  \llangle u, v \rrangle_{\chi} \equiv \< u, {\mathcal{F}}^{-1}(\hat\chi^{-1}\hat v)\>_{L^2}\,,
\end{equation}
where $\mathcal{F}$ denotes the Fourier transform operator and $\hat f
\equiv \mathcal{F}(f)$ and the norm $\|\cdot\|_{\chi}$ is defined
accordingly. The action functional~\eqref{eq:gmam_action}, also termed
the \emph{geometric} action functional, was considered in
\cite{heymann-vandeneijnden:2008, heymann-vandeneijnden:2008b,
  grafke-grauer-schaefer-vandeneijnden:2014} as foundation of a global
minimization technique, the geometric minimum action method (gMAM).

In the context of this work, it is used to derive the corresponding
geometric equations of motion:
\begin{subequations}
  \label{eq:instanton}
  \begin{eqnarray}
    \dot u &=& \frac{\|\dot u\|_{\chi}}{\|b[u]\|_{\chi}}\left(b[u] + \chi\star p\right) \label{eq:instantonu}\\
    \dot p &=& - \frac{\|\dot u\|_{\chi}}{\|b[u]\|_{\chi}} \big( (\nabla b[u])^T p + \lambda \nabla F[u] \delta(t) \big). \label{eq:instantonp}
  \end{eqnarray}
\end{subequations}
With this set of equations of motion it is possible to numerically
discretize the instanton configuration with a finite number of grid
points even if the trajectory is traversed in infinite physical
time. Furthermore, the arc-length parametrization, using the
correlation as metric for the phase space, naturally improves the
numerical resolution in phases with critical dynamics, as it
automatically adapts to the evolution of the action density.

\section{Numerical solution of the instanton equations}
\label{sec:3}

Above, the problem of the computation of instanton configurations has
been narrowed down to numerically finding the solution to the
instanton equations~\eqref{eq:instanton_t} or \eqref{eq:instanton} in
the case of $T\rightarrow\infty$, together with an initial condition
for the field $u(x,-T)=u_\text{start}(x)$ and a final condition for the
auxiliary field of the form $p(x,0)=p_\text{end}(x)=-\lambda \nabla
F[u]$. As the field equation and the auxiliary equation are mutually
dependent, they cannot be solved without knowledge of the solution of
other equation. Therefore, an iterative solution of the instanton
equations was proposed in \cite{chernykh-stepanov:2001}. For any
approximation $u_k(x,t)$ of the field variable, we can solve the
auxiliary equation~\eqref{eq:instantonp} backwards in time, starting
the integration with from the final condition $p_\text{end}(x)$, to
obtain an approximation $p_k(x,t)$. We then use this approximation to
solve the field equation~\eqref{eq:instantonu} forward in time,
starting from its initial condition $u_\text{start}(x)$, to obtain the
approximation $u_{k+1}(x,t)$ for the field variable. The iteration is
started with an initial guess for the field variable $u_0(x,t)$
(e.g. $u_0(x,t)=0$). We iterate until a fixed point is reached, at
which point $(u_k(x,t), p_k(x,t))$ are a solution to the instanton
equations.

First, note that this scheme requires the storage of the field
variable and the auxiliary variable for all times. This issue will be
addressed below. Furthermore, the instanton equations are formally
very similar to the original noisy system~\eqref{eq:spde}. The
numerical solution can therefore be accomplished with a similar set of
tools as the solution of the underlying fluid equation. In particular,
for the field equation it is possible to reuse existing codes that
solve a stochastically forced fluid system and simply replace the
random force with the deterministic term $\chi\star p$. The auxiliary
field, as visible in the example applications below, usually has a
very similar structure. This is in contrast to the variational
approach, which requires the development of a new toolset to
numerically minimize the action functional.

\subsection{Recursive solution of the mixed initial/final value problem}

The most obvious challenge in computing the instanton in the above
laid out fashion for large systems is the requirements in memory. Both
for global minimization techniques and for iterative solutions of the
equations of motion, the field configuration has to be computed
\emph{and saved} for every instance in time, which basically increases
the problem dimension by one. For a class of related problems, an
effective algorithm is proposed in \cite{celani-cencini-noullez:2004}
which scales as $\mathcal{O}(\log N_t)$ in memory consumption instead
of $\mathcal{O}(N_t)$, with only a logarithmic increase in
computational cost. In this section, we present the extension of this
algorithm to the class of mutually dependent initial/final value
problems in the case of instanton equations with large scale force
correlations. Note that similar ideas are known in the field of PDE
constrained optimization as ``checkpointing''
\cite{wang-moin-iaccarino:2009}.

The instanton equations \eqref{eq:instanton} represent a mixed
initial/final value problem: Half of the unknowns are given an initial
value and propagated forward in time, while the others are given a
final value and are propagated backwards in time. These conditions are
also encountered in other physical contexts, e.g. the transport of a
passive scalar by an evolving flow with a given final density, as
described in \cite{celani-cencini-noullez:2004}. Consider a simpler
system of equations of the form
\begin{subequations}
  \label{eq:modelproblem}
  \begin{align}
    u_t = f(u, t), \qquad u(t_1) =& u_1 \label{eq:model1}\\
    p_t = g(u, p, t), \qquad p(t_2) =& p_2\label{eq:model2}
  \end{align}
\end{subequations}
with $t_1 < t_2$, as considered in
\cite{celani-cencini-noullez:2004}. The naive way to solve the system
\eqref{eq:modelproblem} would be to first solve
equation~\eqref{eq:model1} forward in time, starting with the initial
condition $u_1$ at $t_1$, while saving the complete evolution of
$u(t)$ along the way, and subsequently solving
equation~\eqref{eq:model2} backward in time, starting from $p_2$ at
$t_2$, using the previous solution $u(t)$ to evaluate $g(u,p,t)$ along
the way. This approach is $\mathcal{O}(N_t)$ in both memory and
computing time. It represents the worst case in memory usage (every
time-step is saved) and the optimal case in computation time (every
time-step is only computed once). The other extreme would be to only
compute equation~\eqref{eq:model2} backwards in time, starting from
$p_2$ at $t_2$, and computing the needed $u(t)$ at each instance in
time by integrating equation~\eqref{eq:model1} from $t_1$ to
$t$. Here, the memory cost is $\mathcal{O}(1)$, while the
computational cost scales like $\mathcal{O}(N_t^2)$. Of course, this
variant is only possible due to the fact that
equation~\eqref{eq:model1} is independent of the field $p$. In the
case of mutual dependence, such as the system of instanton
equations~\eqref{eq:instanton}, this variant does not apply.

A sensible compromise between memory and computing time can be
achieved when realizing that both above algorithms can be used as
building blocks in the following sense: Divide the interval $[t_1,
  t_2]$ into $k$ sub-intervals. Compute the solution of
equation~\eqref{eq:model1} forward in time from $t_1$ to $t_2$, saving
$u(t)$ along the way only at the start of each of the $k$
intervals. Now, it is possible to solve the original problem on each
sub-interval of size $N_t/k$ with either of the two algorithms from
above. Note that this modification merely changes the constants, but
does not modify the scaling behavior of memory and computing time.

\begin{figure}[tb]
  \begin{center}
    \includegraphics[width=0.6\linewidth]{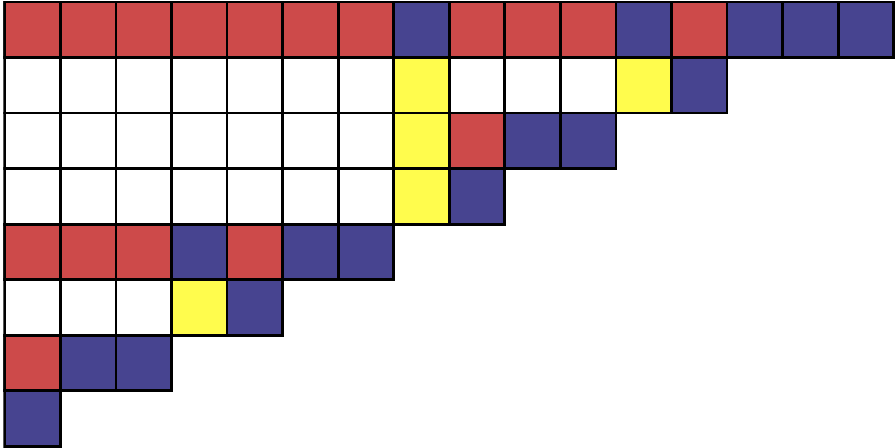}
  \end{center}
  \caption{Depiction of the recursive integration algorithm for
    $k=2$. Each square represents one of the 16 time-steps, showing
    steps computed but subsequently dropped (red), computed and saved
    to memory (blue/dark) and loaded from memory (yellow/light). Total memory
    requirement is the maximum number of green and blue boxes in a
    line (5 in this example), which is $\mathcal{O}(\log N_t)$. Total
    computation time is the total number of red and green boxes (33 in
    this example), which is $\mathcal{O}(N_t\,\log N_t)$.}
  \label{fig:recursive}
\end{figure}

Finally, since for each sub-interval the problem in structure
resembles the original problem, the division outlined above can be
recursively repeated, until the interval size corresponds to the time
discretization $\delta t$. At this point, no further solution inside
the sub-interval is necessary, since the interval represents only one
integration step. As inspired by similar principles in multi-grid
algorithms or the fast Fourier-transform, a natural choice is $k=2$.
Then, the memory requirement scales as $\mathcal{O}(\log N_t)$ and
computing time as $\mathcal{O}(N_t\, \log N_t)$. A schematic depiction
of the algorithm for $k=2$ is shown in Figure~\ref{fig:recursive}. For
the initial solution of the field equation, the field is stored at the
intermediate timesteps $i\in\{8,12,14,15,16\}$, of which the last 3 can
immediately be used for the backwards propagating auxiliary
equation. Whenever a timestep is encountered for which the field
configuration is not stored, it is propagated forward again from the
last known position, while storing intermediate values again in the
fashion lined out above. The choice $k=2$ proves to be a good
compromise between memory and computing time, even though any $k$ is
admissible to use the available memory at optimal efficiency (even
fractional $k$ are possible, choosing the interval boundaries at $t_i
= t_1 + \frac{t_2 - t_1}{k}i$, $i \in \{1, \dots, \lfloor k
\rfloor\}$). In the following section, we will apply this algorithm to
instanton equations of the form~\eqref{eq:instanton} by taking
advantage of the form of the forcing
correlation~\eqref{eq:space_corr}.


\subsection{Projecting the auxiliary field}

In fact, the instanton equations~\eqref{eq:instanton} are not of the
form of equation~\eqref{eq:modelproblem}:
Equations~\eqref{eq:instantonu} and \eqref{eq:instantonp} mutually
depend on each other, prohibiting the purely recursive form lined out
above, because neither $u$ nor $p$ can be integrated without knowledge
of the other field. It is therefore necessary to store at least one
field completely. Yet, the dependence of the $u$-equation on the $p$
is solely through the term $\chi \star p$, which, due to its compact
support in Fourier space (equation~\eqref{eq:space_corr}), acts only
on few modes. The convolution of the auxiliary field with the forcing
correlation can therefore be seen as a projection on the active modes
of the forcing.

Consequently it is advantageous, instead of saving any of the field
configurations $u$ or $p$, to only store the projection, i.e. the
non-vanishing modes of the field $\chi \star p$ for all time-steps.
From this one can reconstruct the velocity field $u$ in the recursive
manner lined out above. This approach does not recover the
$\mathcal{O}(\log N_t)$ scaling in memory, but restricts the number of
degrees of freedom in every other dimension, therefore preserving
advantageous memory requirements. More importantly, since the physical
cut-off scale $k_c$ is a constant, the memory cost of the auxiliary
field becomes independent of the number of degrees of freedom in space
$N_x$.

Also note that in some applications it is actually feasible to
integrate the whole auxiliary equation~\eqref{eq:instantonp} in the
convoluted field variable instead. Even though this approach does not
reduces the number of active modes (as mixing of modes is possible
through the nonlinear drift term), it provides a beneficial
mollification of the auxiliary variable due to the large-scale nature
of the correlation function $\chi(x)$.

\section{Applications}
\label{sec:4}

We now present several applications of the proposed algorithm. Every
integration of the instanton equations~\eqref{eq:instanton} is done
with a second order Runge-Kutta integrator in time. Spatial
derivatives are computed with a pseudo-spectral scheme using fast
Fourier transforms (FFT). Due to the high memory efficiency of the
presented algorithm a speedup of the computation using graphic cards
becomes feasible. All computations presented hereafter where conducted
on a single graphics card using the CUDA toolkit \cite{cuda:2014}.

We want to focus on the Burgers equation,
\begin{equation}
  \label{eq:burgers}
  \partial_t \mathbf{u} + \mathbf{u}\cdot\nabla\mathbf{u} - \nu \Delta\mathbf{u} = \mathbf{f} 
\end{equation}
in one or two dimensions, $\mathbf{u}(x,y,t): \mathbb{R}^d
\times [-T,0] \rightarrow \mathbb{R}^d$, $d=1$ or $d=2$, with force
correlation
\begin{equation}
  \langle f_i(\mathbf{x}+\mathbf{r},s+t) f_j(\mathbf{x},s) \rangle = \delta(t) \chi_{ij}(r)
\end{equation}
of the form~\eqref{eq:correlation}. For this system we want to solve
the instanton equations~\eqref{eq:instanton} numerically. As
observable we choose the gradient of the velocity field $\mathbf{u}$
in the origin for $d=1$, or a sensible generalization thereof for
$d=2$, in order to find the most probable evolution of an extreme
gradient event out of a stationary turbulent Burgers flow.

\subsection{One dimensional Burgers equation}

For the 1D case, let the correlation in space $\chi(r)$ be of the
form~\eqref{eq:space_corr} with the shape being a Mexican hat, $f(k) =
-k^2 \exp(-k^2/2)$. As observable we choose
\begin{equation}
  \label{eq:observable-1d}
  F[u] = \partial_x u(0,0)\,,
\end{equation}
which conditions on the gradient of the velocity field in the
space-time origin. This amounts to finding the most probable evolution
of the 1D Burgers equation towards a shock with gradient $a$ at $x=0$
as a final configuration.

\begin{figure}[tb]
  \begin{center}
    \includegraphics[width=0.8\linewidth]{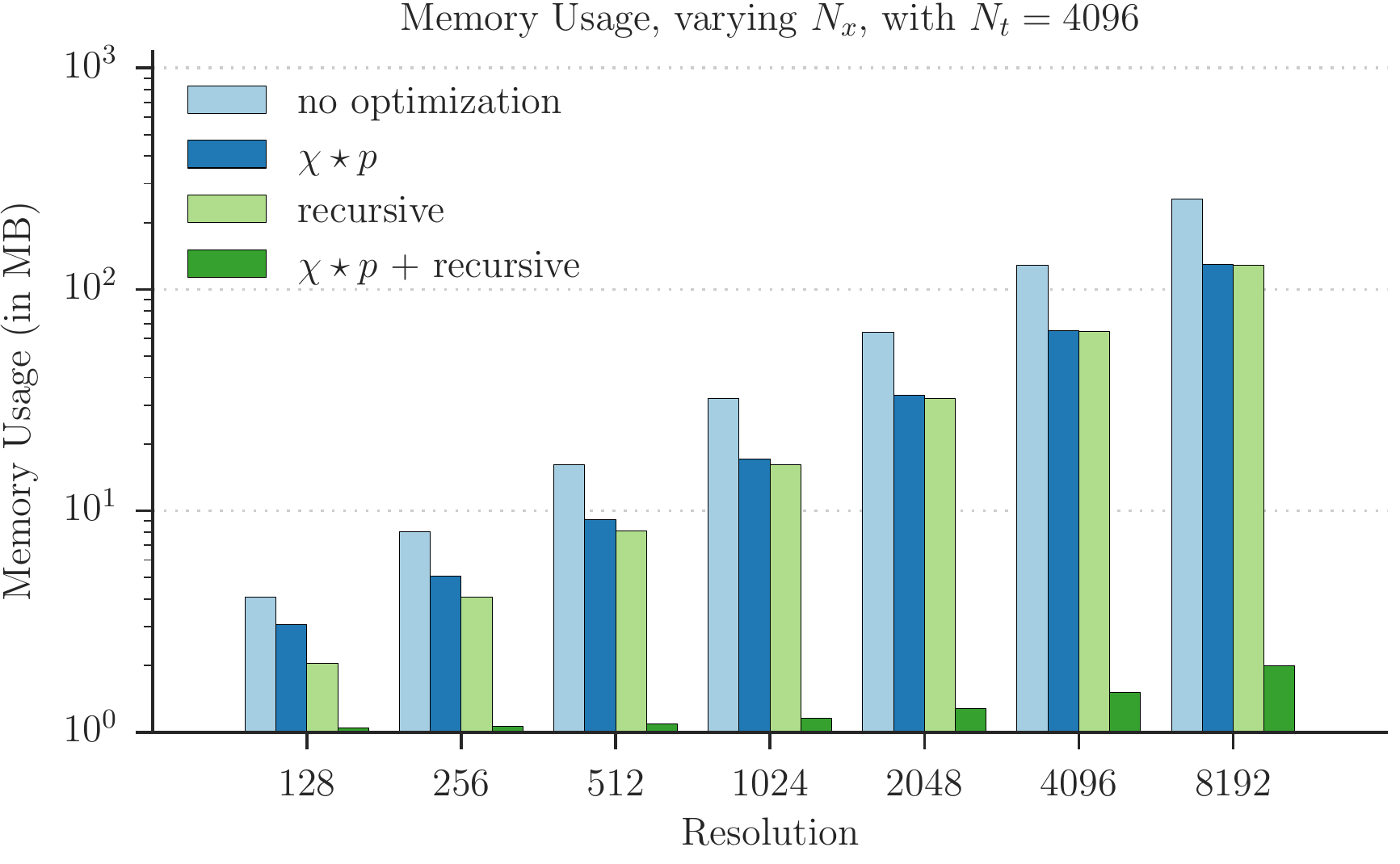}
  \end{center}
  \caption{Comparison of memory costs for 1D simulations with varying
    spatial resolution and a fixed time resolution of
    $N_t=4096$. Using only either the recursive algorithm or the
    compactness of the forcing (denoted as ``$\chi\star p$'') roughly
    halves the memory usage. Combining both methods leads to an
    improvement of memory efficiency of more than a factor $10^2$
    ($257$MB naive vs.~$2$MB optimized).
  \label{fig:memuse1d}}
\end{figure}

The physical implications of this choice are discussed in
\cite{grafke-grauer-etal:2014}. In particular, the
computation of the instanton for the
observable~\eqref{eq:observable-1d} allows the computation of the
shape of the far left tail for the velocity gradient probability
distribution function. In addition to being a computational tool, the
instanton configuration was furthermore demonstrated to correspond to
the most extreme events in turbulent Burgers flows
\cite{grafke-grauer-schaefer:2013}. In particular, it was shown that
the instanton configuration can be extracted directly from a
stochastically driven Burgers flow, and corresponds in shape and
evolution to the prediction.

Here, we focus on the computational efficiency of the proposed
algorithm. For memory usage statistics, see
figure~\ref{fig:memuse1d}. Note that both the projection of the
forcing onto active modes and the recursive time integration each have
only little effect on the memory consumption on their own. Only the
combination of both optimizations leads to a drastic reduction in
memory usage of about a factor~100. In particular, for the highest
resolution that was tested, only 2MB of memory were required, in
comparison to 257MB of the unoptimized version. On the other hand, due
to the performance scaling with $\mathcal{O}(N_t \log N_t)$, the
optimizations impose a computational cost of merely a factor~3
compared to the algorithm without any optimizations, which is small in
comparison to the achieved memory savings.

\subsection{Two dimensional Burgers equation}

\begin{figure}[tb]
  \begin{center}
    \includegraphics[width=0.48\linewidth]{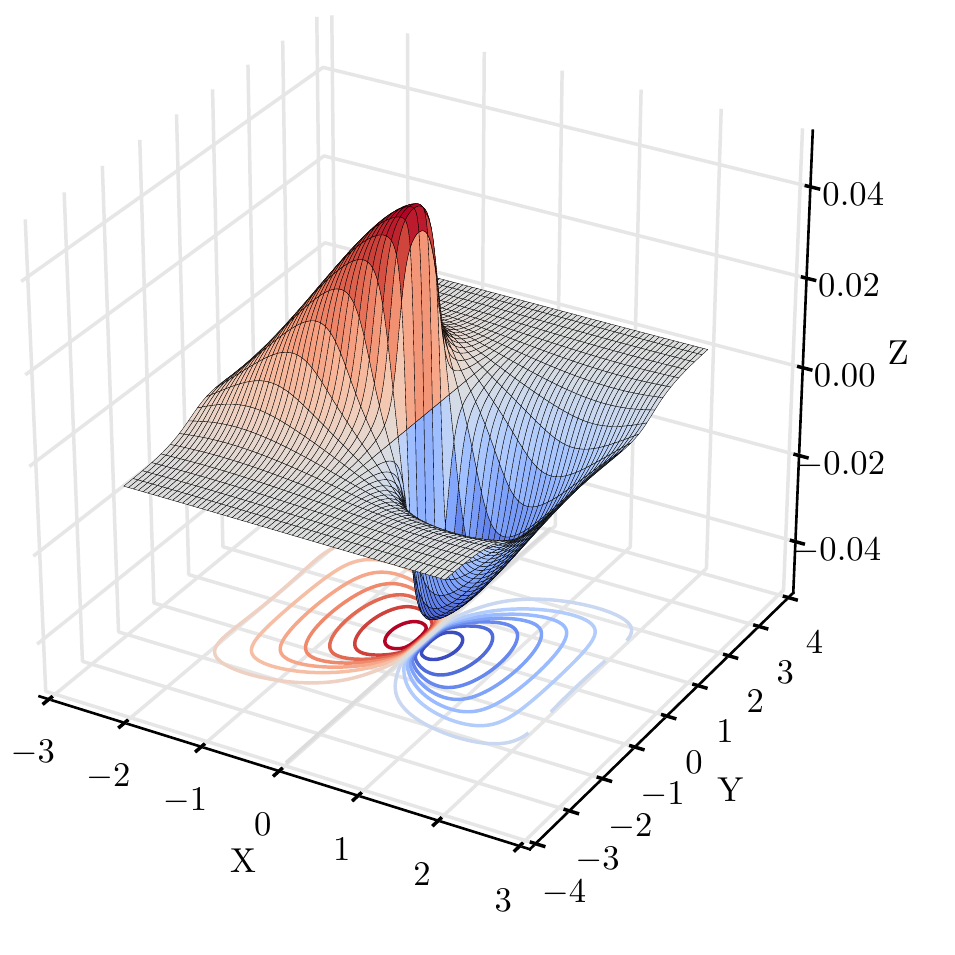}
    \includegraphics[width=0.48\linewidth]{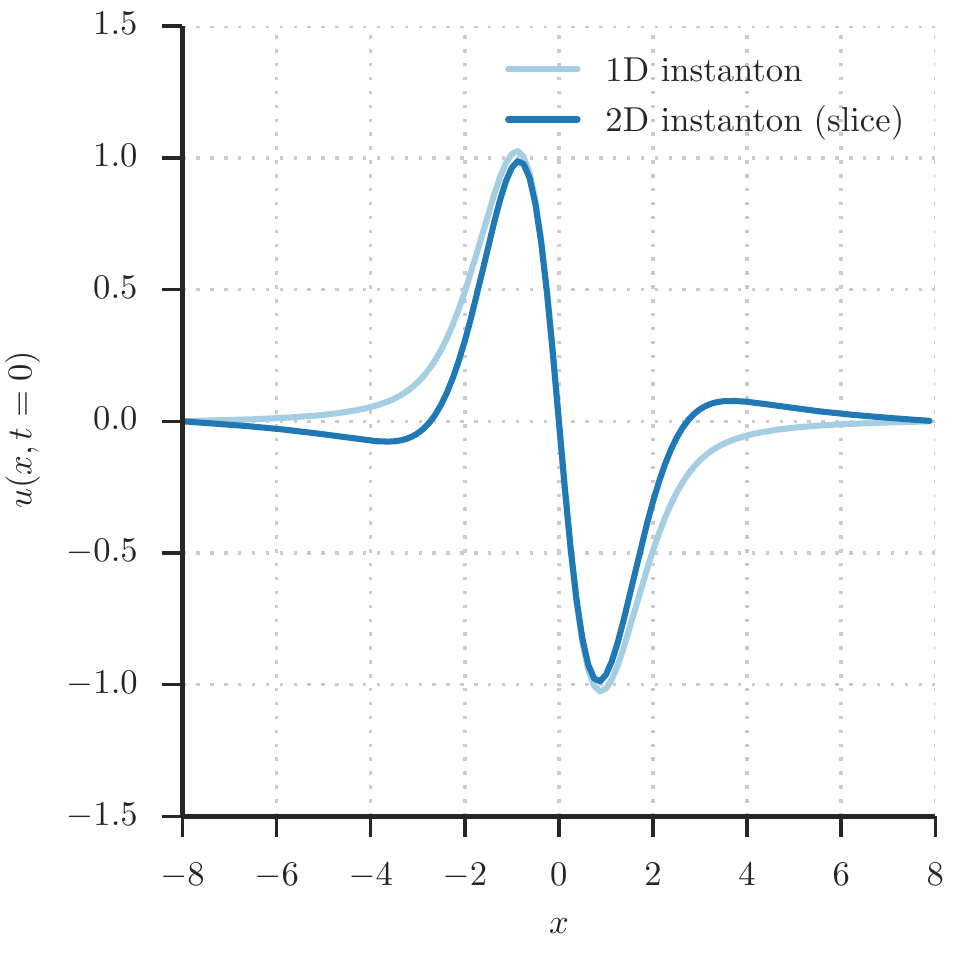}
  \end{center}
  \caption{Left: Surface plot of the $x$-component of the velocity
    field, $u_x(x,y)$, for the 2D shock instanton configuration for
    the gradient $\partial_x u_x=-1$, $\nu=10^{-2}$. Right: Comparison
    between the 1D shock instanton configuration and a slice through
    the 2D shock instanton configuration for the gradient $\partial_x
    u_x = -2$, $\nu=\tfrac12$.}
  \label{fig:2dinstanton-b}
\end{figure}

\begin{figure}[tb]
  \begin{center}
    \includegraphics[width=\linewidth]{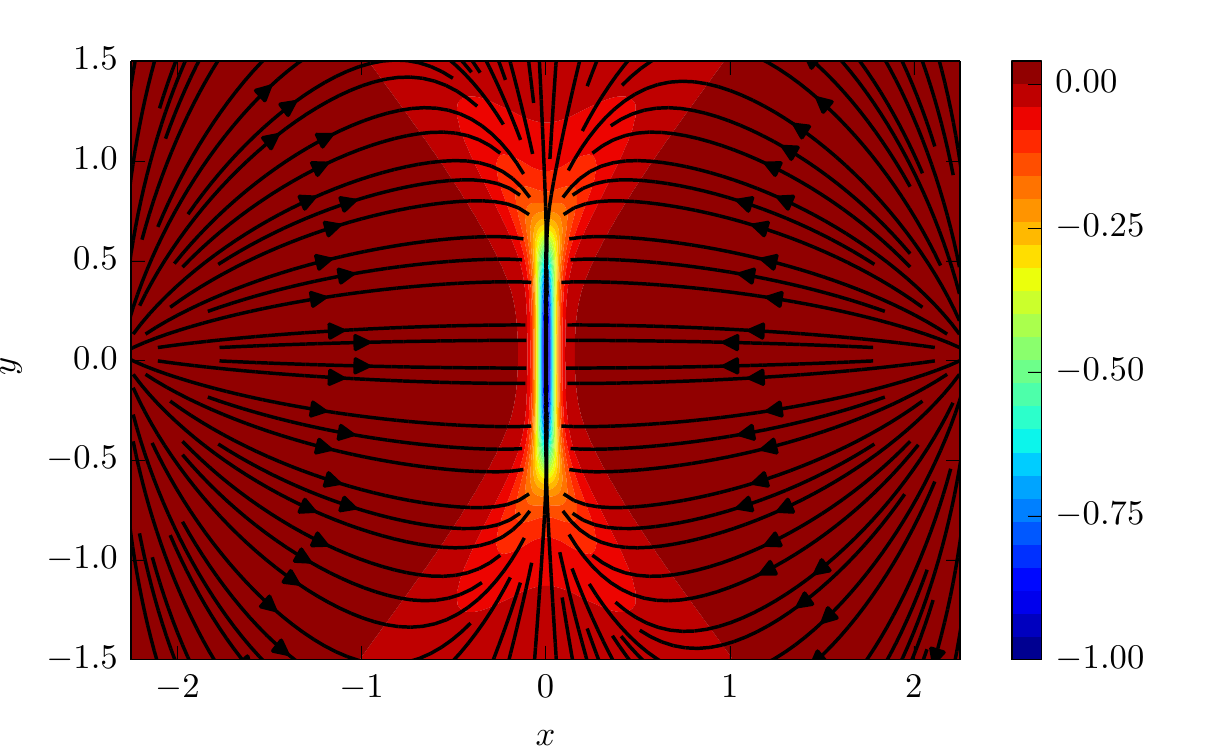}
  \end{center}
  \caption{Contour plot of the velocity gradient in $x$-direction,
    $\partial_x u_x$ for the 2D instanton configuration for a gradient
    $\partial_x u_x = -1$ at $t=0$. The arrows depict the direction of
    the velocity field $u(x,y)$.}
  \label{fig:2dinstanton}
\end{figure}

\begin{figure}[tb]
  \begin{center}
    \includegraphics[width=0.48\linewidth]{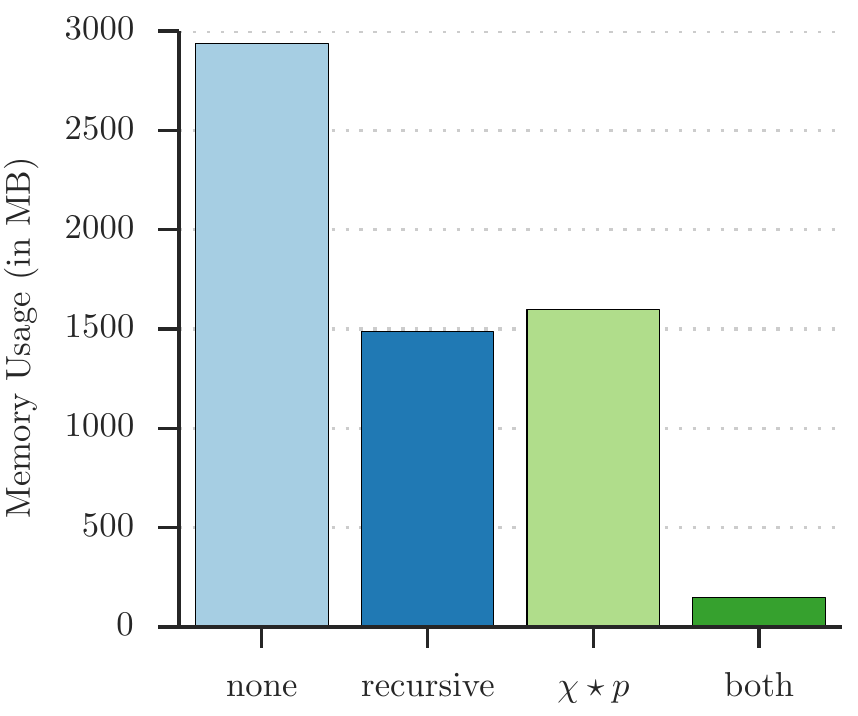}
    \includegraphics[width=0.48\linewidth]{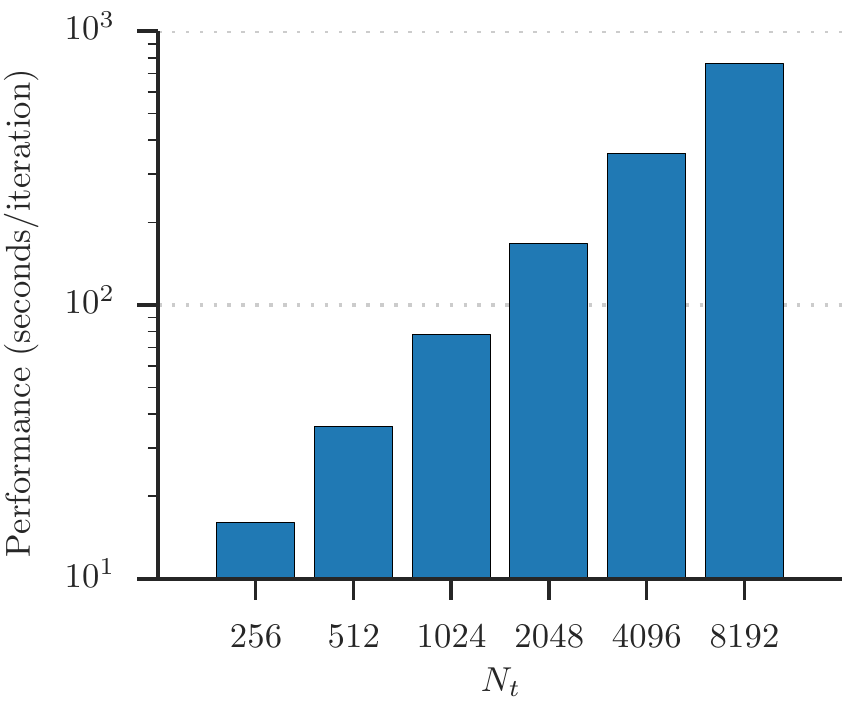}
  \end{center}
  \caption{Left: Comparison of memory costs for 2D simulations with
    $N_x=256\times256$, $N_t=2048$ for the presented
    optimizations. The total memory saving of the combined algorithm
    exceeds a factor of $20$. Right: Performance of the optimized
    algorithm for $N_x=1024\times1024$ and varying $N_t$ scales as
    $\mathcal{O}(N_t \log N_t)$.}
  \label{fig:2dinstanton-memory}
\end{figure}

The 2D Burgers equation preserves irrotationality of the flow under
corresponding forcing. Therefore, it is natural to decompose the
forcing correlation function according to
\begin{equation}
  \chi_{ij}(r) = \alpha \chi_{ij}^\text{irr}(r) + (1-\alpha) \chi_{ij}^\text{sol}(r)\,,
\end{equation}
where $\chi_{ij}^\text{irr}$ and $\chi_{ij}^\text{sol}$ denote
irrotational and solenoidal forcing respectively. In two dimensions,
they are realized through
\begin{eqnarray}
  \chi_{ij}^\text{irr}(r) &=& g(r) \delta_{ij} + r g'(r) \frac{r_i r_j}{r^2} \\
  \chi_{ij}^\text{sol}(r) &=& f(r) \delta_{ij} + \frac{r f'(r)}{d-1} \left(\delta_{ij}-\frac{r_i r_j}{r^2}\right)\,.
\end{eqnarray}
Note that the irrotational forcing is independent of the spatial
dimension $d$, whereas the solenoidal forcing explicitly contains
$d-1$ in the denominator of the second term.

In one dimension, only the irrotational forcing is allowed and reduces
to
\begin{equation*}
  \chi_{11}^\text{irr}(r) = g(r) + r g'(r)\,.
\end{equation*}
In the relevant case of large scale forcing, we require up to the
cut-off scale that
\begin{equation}
  g(r) + r g'(r) = -\partial_{rr} e^{-\frac{r^2}{2}}\,,
\end{equation}
to match the 1D case, which directly leads to
\begin{equation}
  g(r) = e^{-\frac{r^2}{2}} \; .
\end{equation}
We will restrict ourselves to purely irrotational forcing, $\alpha=1$,
in the following. The instanton equations~\eqref{eq:instanton_t} for
the 2D Burgers equation read
\begin{subequations}
  \label{eq:2d-instantons}
  \begin{align}
    \partial_t \mathbf{u} + \mathbf{u} \cdot \nabla \mathbf{u} - \nu \Delta \mathbf{u} &= \chi \star \mathbf{p}\\
    \partial_t \mathbf{p} + \mathbf{u} \cdot \nabla \mathbf{p} - (\mathbf{p} \times \nabla) \mathbf{u}^\perp + \nu \Delta \mathbf{p} &= 0\,,
  \end{align}
\end{subequations}
with $\mathbf{u}^\perp = (-u_y, u_x)$ and $(\chi \star \mathbf{p})_i =
\sum_j \chi_{ij} \star p_j$. There are several choices to generalize
the observable \eqref{eq:observable-1d} to more than one space
dimension: Setting $F[\mathbf{u}] = \nabla \cdot \mathbf{u}(0,0)$
selects configurations of extreme velocity divergence in the origin,
which amounts to compressible ``explosions'' for $a>0$ and
``implosions'' for $a<0$. Here, we will focus on the choice
$F[\mathbf{u}] = \partial_x u_x(0,0)$. This choice corresponds to high
gradients in $x$-direction, resulting in prototypical 2D shock
configurations that dominate the statistics of turbulent 2D Burgers
flows.

The shock structure for a gradient $\partial_x u_x(0,0)=-1$ is
depicted in the surface plot of the $x$-component of the velocity
field $\mathbf{u}(x,y)$ at the final time $t=0$ in
figure~\ref{fig:2dinstanton-b} (left). This configuration constitutes
the most probable shock structure with the chosen gradient in the
limit of small forcing. A comparison between the 1D instanton
configuration and a cut through the 2D instanton configuration
perpendicular to the shock at $y=0$ is presented in
figure~\ref{fig:2dinstanton-b} (right). Note that the
higher-dimensional setup cannot reduce to a completely one-dimensional
structure independent of $y$, since the assumption of isotropy of the
forcing, in combination with the optimal forcing in $x$-direction,
leads to a residual non-vanishing velocity component $u_y$. The size
of the shock structure along the shock is therefore determined by the
correlation length of the forcing $L=1$, while its size perpendicular
to the shock is determined by the viscosity $\nu \ll 1$. This is
clearly visible in figure~\ref{fig:2dinstanton}, where the contour
plot shows the $x$-component of the gradient in $x$-direction,
$\partial_x u_x$, and the arrows depict the direction of the velocity
field $\mathbf{u}(x,y)$ with $N_x=2048\times512$, $N_t=2048$ and
$\nu=0.01$.

The highest resolution achieved is $N_x=1024\times1024$, $N_t=8192$,
with a total memory usage of $577$MB, performance of $764
\text{s}/\text{iter}$ on a desktop computer, Intel Xeon E5-1620
(3.6GHz), GeForce GTX680 (GK104 ``Kepler''). This amounts to a number
of degrees of freedom of $N=N_x\times N_t=2^{33}\approx10^{10}$ on a
single graphics card, which is about a factor $10^5$ more than
state-of-the-art global minimization algorithms in 2D settings
(e.g. nucleation in the presence of shear with $N_x=64^2$, $N_t=100$
\cite{heymann-vandeneijnden:2008c}, minimizers of the KPZ-equation
with $N_x=100^2$, $N_t=100$ \cite{fogedby-ren:2009}, or geophysical
bi-stability for the quasi-geostrophic equations, $N_x=16^2$,
$N_t=200$ \cite{laurie-bouchet:2014}). The extrapolated memory usage
of the unoptimized algorithm is $183$GB, which would exceed our
implementation by more than a factor $300$, while the computational
overhead of the combined optimization is again below a mere factor $3$
in the total computation time. As the scaling in memory is sub-linear,
it is to be expected that state-of-the-art hardware will allow for a
disproportionally higher number of degrees of freedom to be computed.
Figure~\ref{fig:2dinstanton-memory} (left) shows a comparison of the
memory cost for a lower resolution setup of $N_x=256\times256$,
$N_t=2048$ in order to fit the unoptimized case on the machine: Both
the recursive optimization and the projection method amount to a
saving of roughly a half, the combination of both leads to memory
savings of a factor $20$. On the other hand, the computational
performance overhead of the optimizations scales with $N_t$ as
$\mathcal{O}(N_t \log N_t)$ even for large spatial resolution. This is
depicted in figure~\ref{fig:2dinstanton-memory} (right) for the fully
optimized algorithm for $N_x = 1024\times1024$ and varying
$N_t$. Because of memory restrictions, a comparison to the unoptimized
variant is only possible for the lowest value of $N_t=256$ with
$7.9~\text{s}/\text{iter}$ (unoptimized memory usage) versus
$16~\text{s}/\text{iter}$ (optimized memory usage).

\section{Conclusion and discussion}
\label{sec:5}

We present a novel method for computing minimizers of the action
functional arising in the context of the instanton formalism for the
computation of extreme events for stochastic partial differential
equations. In particular, it is applicable in the setup of
multi-dimensional fluid equations with a direct energy cascade, where
the energy is injected on large scales.

The algorithm combines several ideas from other fields to optimize the
memory footprint of the minimization problem at a high number of
degrees of freedom in both temporal and spatial direction at marginal
computational overhead: In the spirit of
\cite{chernykh-stepanov:2001}, mixed initial/final boundary conditions
of the equations of motion are integrated in time instead of solving
the global optimization problem. Disadvantageous time-discretization
in the presence of an infinite time minimizer is mitigated by
employing the geometric action functional
\cite{heymann-vandeneijnden:2008,
  grafke-grauer-schaefer-vandeneijnden:2014} instead of the classical
Martin-Siggia-Rose/Janssen/de Dominicis response functional. In time
direction, we then extent a multigrid-inspired recursive time
integration technique \cite{celani-cencini-noullez:2004} to the
setting of mutually dependent instanton equations. This is only
effective in combination with a projection of the auxiliary field on
its active modes, which is made possible by the large-scale
correlation in space of the driving force in the turbulent
regime. These optimizations amount to a $\mathcal{O}(\log N_t)$
dependence in memory of the field variable, while the memory usage of
the auxiliary field becomes decoupled from the spatial resolution
$N_x$. The computational effectiveness of the algorithm is demonstrated
in the case of one- and two-dimensional compressible turbulence for
the Burgers equation. We determine and compare the instanton
configuration for a gradient observable in the 1D and 2D case, which
amounts to computing the prototypical shock evolution and final
configuration with the imposed gradient value.

Additionally, we demonstrate the ability to compute minimizing
trajectories with as much as $N\approx 10^{10}$ degrees of freedom.
The algorithm is therefore suitable to numerically determine the
instanton field configuration for two- and three-dimensional
incompressible turbulence that is conjectured to have significant
impact on the evolution of turbulent fluids and dominate the tails of
turbulent statistics.

\section*{Acknowledgments}

The work of T.G. was partially supported through the grants
ISF-7101800401 and Minerva – Coop 7114170101. R.G.and S.S. acknowledge
support through DFG-FOR1048. T.G. thanks Gregory Falkovich for the
support. The authors acknowledge Patrick Teubner for the
implementation of wavelet compression, and thank Tobias Sch\"afer for
helpful discussions.


\begin{thebibliography}{10}

\bibitem{cuda:2014}
{\em CUDA toolkit documentation}, 2014.
\newblock http://docs.nvidia.com/cuda/index.html.

\bibitem{assaf-roberts-etal:2011}
Michael Assaf, Elijah Roberts, and Zaida Luthey-Schulten.
\newblock Determining the stability of genetic switches: Explicitly accounting
  for mrna noise.
\newblock {\em Phys. Rev. Lett.}, 106:248102, 2011.

\bibitem{backus-chernov-martin:2011}
David Backus, Mikhail Chernov, and Ian Martin.
\newblock Disasters implied by equity index options.
\newblock {\em The Journal of Finance}, 66:1969--2012, 2011.

\bibitem{balkovsky-falkovich-etal:1997}
E.~Balkovsky, G.~Falkovich, I.~Kolokolov, and V.~Lebedev.
\newblock Intermittency of {B}urgers' turbulence.
\newblock {\em Phys. Rev. Lett.}, 78:1452, 1997.

\bibitem{barro:2006}
Robert~J. Barro.
\newblock Rare disasters and asset markets in the twentieth century.
\newblock {\em Quarterly Journal of Economics}, 2006.

\bibitem{celani-cencini-noullez:2004}
Antonio Celani, Massimo Cencini, and Alain Noullez.
\newblock Going forth and back in time: a fast and parsimonious algorithm for
  mixed initial/final-value problems.
\newblock {\em Physica D: Nonlinear Phenomena}, 195(3):283--291, 2004.

\bibitem{chernykh-stepanov:2001}
A.~I. Chernykh and M.~G. Stepanov.
\newblock Large negative velocity gradients in {B}urgers turbulence.
\newblock {\em Phys. Rev. E}, 64:026306, 2001.

\bibitem{dedominicis:1976}
C.~de~Dominicis.
\newblock Techniques de renormalisation de la th\'{e}orie des champs et
  dynamique des ph\'enom\`enes critiques.
\newblock {\em J. Phys. C}, 1:247, 1976.

\bibitem{dembo-zeitouni:2010}
Amir Dembo and Ofer Zeitouni.
\newblock {\em Large deviations techniques and applications}.
\newblock Springer-Verlag, Berlin, 2010.

\bibitem{e-ren-vandeneijnden:2004}
W.~E, W.~Ren, and E.~Vanden-Eijnden.
\newblock Minimum action method for the study of rare events.
\newblock {\em Commun. Pure Appl. Math.}, 57:1--20, 2004.

\bibitem{e-ren-vandeneijnden:2002}
Weinan E, Weiqing Ren, and Eric Vanden-Eijnden.
\newblock String method for the study of rare events.
\newblock {\em Phys. Rev. B}, 66(5):052301, 2002.

\bibitem{elgart-kamenev:2006}
Vlad Elgart and Alex Kamenev.
\newblock Classification of phase transitions in reaction-diffusion models.
\newblock {\em Phys. Rev. E}, 74:041101, 2006.

\bibitem{fogedby-ren:2009}
H.~C. Fogedby and W.~Ren.
\newblock Minimum action method for the {K}ardar-{P}arisi-{Z}hang equation.
\newblock {\em Phys. Rev. E}, 80:041116, 2009.

\bibitem{freidlin-wentzell:2012}
Mark~I Freidlin and Alexander~D Wentzell.
\newblock {\em Random perturbations of dynamical systems}, volume 260.
\newblock Springer, 2012.

\bibitem{grafke-grauer-schaefer:2013}
T.~Grafke, R.~Grauer, and T.~Sch\"afer.
\newblock Instanton filtering for the stochastic {B}urgers equation.
\newblock {\em J. Phys. A}, 46(6):62002, 2013.

\bibitem{grafke-grauer-etal:2014}
T.~Grafke, R.~Grauer, T.~Sch\"afer, and E.~Vanden-Eijnden.
\newblock Relevance of instantons in {B}urgers turbulence.
\newblock {\em arXiv preprint arXiv:1412.0255}, December 2014.

\bibitem{grafke-grauer-schaefer-vandeneijnden:2014}
T.~Grafke, R.~Grauer, T.~Sch\"afer, and Eric Vanden-Eijnden.
\newblock Arclength parametrized {H}amilton's equations for the calculation of
  instantons.
\newblock {\em Multiscale Modeling \& Simulation}, 12(2):566--580, 2014.

\bibitem{gurarie-migdal:1996}
V.~Gurarie and A.~Migdal.
\newblock Instantons in the {B}urgers equation.
\newblock {\em Phys. Rev. E}, 54:4908, 1996.

\bibitem{henkelman-jonsson:2000}
Graeme Henkelman and Hannes J{\'o}nsson.
\newblock Improved tangent estimate in the nudged elastic band method for
  finding minimum energy paths and saddle points.
\newblock {\em The Journal of Chemical Physics}, 113:9978, 2000.

\bibitem{heymann-vandeneijnden:2008c}
M.~Heymann and E.~Vanden-Eijnden.
\newblock Pathways of maximum likelihood for rare events in nonequilibrium
  systems: application to nucleation in the presence of shear.
\newblock {\em Phys. Rev. Lett.}, 100(14):140601, 2008.

\bibitem{heymann-vandeneijnden:2008}
M.~Heymann and E.~Vanden-Ejnden.
\newblock The geometric minimum action method: A least action principle on the
  space of curves.
\newblock {\em Commun. Pure Appl. Math.}, 61:1053, 2008.

\bibitem{janssen:1976}
H.K. Janssen.
\newblock On a {L}agrangian for classical field dynamics and renormalization
  group calculations of dynamical critical properties.
\newblock {\em Z. Physik B}, 23:377, 1976.

\bibitem{kramers:1940}
H.~A. Kramers.
\newblock Brownian motion in a field of force and the diffusion model of
  chemical reactions.
\newblock {\em Physica}, 7(4):284--304, April 1940.

\bibitem{laurie-bouchet:2014}
Jason Laurie and Freddy Bouchet.
\newblock Computation of rare transitions in the barotropic quasi-geostrophic
  equations.
\newblock {\em arXiv preprint arXiv:1409.3219}, 2014.

\bibitem{martin-siggia-rose:1973}
P.~C. Martin, E.~D. Siggia, and H.~A. Rose.
\newblock Statistical dynamics of classical systems.
\newblock {\em Phys. Rev. A}, 8:423, 1973.

\bibitem{onorato-residor-etal:2013}
M.~Onorato, S.~Residori, U.~Bortolozzo, A.~Montina, and F.T. Arecchi.
\newblock Rogue waves and their generating mechanisms in different physical
  contexts.
\newblock {\em Physics Reports}, 528(2):47 -- 89, 2013.

\bibitem{schmeits-djikstra:2001}
Maurice~J. Schmeits and Henk~A. Dijkstra.
\newblock Bimodal behavior of the {K}uroshio and the {G}ulf {S}tream.
\newblock {\em J. Phys. Ocean.}, 31:3435, 2001.

\bibitem{vandeneijnden-heymann:2008}
Eric Vanden-Eijnden and Matthias Heymann.
\newblock The geometric minimum action method for computing minimum energy
  paths.
\newblock {\em The Journal of chemical physics}, 128:061103, 2008.

\bibitem{heymann-vandeneijnden:2008b}
E.~Vanden-Ejnden and M.~Heymann.
\newblock The geometric minimum action method for computing minimum energy
  paths.
\newblock {\em Jour. Chem. Phys.}, 128:061103, 2008.

\bibitem{wang-moin-iaccarino:2009}
Q.~Wang, P.~Moin, and G.~Iaccarino.
\newblock Minimal {Repetition} {Dynamic} {Checkpointing} {Algorithm} for
  {Unsteady} {Adjoint} {Calculation}.
\newblock {\em SIAM Journal on Scientific Computing}, 31(4):2549--2567, January
  2009.

\bibitem{zhou-ren-e:2008}
Xiang Zhou, Weiqing Ren, and Weinan E.
\newblock Adaptive minimum action method for the study of rare events.
\newblock {\em J. Chem. Phys.}, 128:104111, 2008.

\end{thebibliography}
\end{document}